\documentclass[pra,twocolumn,showpacs,eqsecnum,superscriptaddress]{revtex4}

\usepackage{graphicx}
\usepackage{amsmath}
\usepackage{verbatim}
\usepackage{color}

\begin{document}
\vfuzz2pt 
\hfuzz2pt 

\newcommand{\sch}{Schr\"odinger}
\newcommand{\schs}{Schr\"odinger's}
\newcommand{\nn}{\nonumber}
\newcommand{\nl}{\nn \\ &&}
\newcommand{\dg}{^\dagger}
\newcommand{\rt}[1]{\sqrt{#1}\,}
\newcommand{\bra}[1]{\langle{#1}|}
\newcommand{\ket}[1]{|{#1}\rangle}
\newcommand{\ito}{It\^o }
\newcommand{\str}{Stratonovich }
\newcommand{\erf}[1]{Eq.~(\ref{#1})}
\newcommand{\erfs}[2]{Eqs.~(\ref{#1}) and (\ref{#2})}
\newcommand{\erft}[2]{Eqs.~(\ref{#1}) -- (\ref{#2})}

\newcommand{\eq}[1]{Eq.~(\ref{#1})}
\newcommand{\eqq}[1]{(\ref{#1})}

\newcommand{\sz}{\hat\sigma_{z}}
\newcommand{\sx}{\hat\sigma_{x}}
\newcommand{\sy}{\hat\sigma_{y}}
\newcommand{\spp}{\hat\sigma_{+}}
\newcommand{\smm}{\hat\sigma_{-}}

\newcommand{\aop}{\hat a}
\newcommand{\ad}{\hat a^\dag}
\newcommand{\veps}{\varepsilon}
\newcommand{\rf}{\mathrm{rf}}
\newcommand{\spec}{\mathrm{s}}

\newcommand{\wrr}{\omega_{r}}
\newcommand{\wa}{\omega_{a}}
\newcommand{\ej}{E_\mathrm{J}}
\newcommand{\ec}{E_\mathrm{C}}
\newcommand{\vlc}{V_\mathrm{LC}}

\newcommand{\ncrit}{n_\mathrm{crit}}

\newcommand{\Gm}{\Gamma_\mathrm{m}}
\newcommand{\tGm}{\tilde\Gamma_\mathrm{m}}

\def\be{\begin{equation}}
\def\ee{\end{equation}}

\title{Qubit-photon interactions in a cavity:\\
Measurement induced dephasing and number splitting}

\date{\today}
\author{Jay~Gambetta}
\affiliation{Departments of Applied Physics and Physics, Yale
University, New Haven, CT 06520}
\author{Alexandre~Blais}
\affiliation{Departments of Applied Physics and Physics, Yale
University, New Haven, CT 06520} \affiliation{D\'epartement de
Physique et Regroupement Qu\'eb\'ecois sur les Mat\'eriaux de
Pointe, Universit\'e de Sherbrooke, Sherbrooke, Qu\'ebec, Canada,
J1K 2R1}
\author{D.~I.~Schuster}
\affiliation{Departments of Applied Physics and Physics, Yale
University, New Haven, CT 06520}
\author{A.~Wallraff}
\affiliation{Departments of Applied Physics and Physics, Yale
University, New Haven, CT 06520} \affiliation{Department of Physics,
ETH Zurich, CH-8093 Z{\"u}rich, Switzerland}
\author{L.~Frunzio}
\affiliation{Departments of Applied Physics and Physics, Yale
University, New Haven, CT 06520}
\author{J.~Majer}
\affiliation{Departments of Applied Physics and Physics, Yale
University, New Haven, CT 06520}
\author{M. H. Devoret}
\affiliation{Departments of Applied Physics and Physics, Yale
University, New Haven, CT 06520}
\author{S.~M.~Girvin}
\affiliation{Departments of Applied Physics and Physics, Yale
University, New Haven, CT 06520}
\author{R.~J.~Schoelkopf}
\affiliation{Departments of Applied Physics and Physics, Yale
University, New Haven, CT 06520}

\begin{abstract}
We theoretically study measurement induced-dephasing of a
superconducting qubit in the circuit QED architecture and compare
the results to those obtained experimentally by Schuster {\it et
al.}, [Phys. Rev. Lett. 94, 123602 (2005)].  Strong coupling of the
qubit to the resonator leads to a significant ac-Stark shift of the
qubit transition frequency.  As a result, quantum fluctuations in
the photon number populating the resonator cause dephasing of the
qubit.   We find good agreement between the predicted line shape of
the qubit spectrum and the experimental results. Furthermore, in the
strong dispersive limit, where the Stark shift per photon is large
compared to the cavity decay rate and the qubit linewidth,
we predict that the qubit spectrum will be split into multiple
peaks, with each peak corresponding to a different number of photons
in the cavity.
\end{abstract}


\maketitle

\section{Introduction}

Superconducting qubits are promising building blocks for the
realization of a quantum computer~\cite{devoret:2004}.   Several
experiments have shown coherent control of a single
qubit~\cite{wallraff:2005,chiorescu:2003,vion:2002,martinis:2002,nakamura:99}
and two-qubit experiments have been
realized~\cite{mcdermott:2005,majer:2005,Yamamoto:2003,Berkley:2003,pashkin:2002}.
Recently, it was suggested that superconducting qubits can be
strongly coupled to distributed or discrete LC circuits in a way
that opens the possibility to study quantum optics related phenomena
in solid-state
devices~\cite{makhlin:2001,blais:2004,blais:2003,plastina:2003,you:2003b}.
This concept has been successfully demonstrated
experimentally~\cite{wallraff:2004,Chiorescu:2004,schuster:2005,wallraff:2005,xu:2005}
and effects associated with the quantum nature of the microwave
electromagnetic field have now been seen in the form of vacuum Rabi
splitting~\cite{wallraff:2004} and measurement induced dephasing via
photon shot noise ~\cite{schuster:2005}.  In this paper we present a
detailed analysis of the quantum fluctuations of the photon number
in the cavity and its effect on the qubit spectrum. We also show
that access to the extreme limit of strong dispersive coupling
should allow direct observation of the photon number distribution in
the cavity.

An advantage of some of these circuit QED analogs of cavity QED
is that the cavity
presents a well defined electromagnetic environment to the qubit
which can lead to enhanced coherence times of the
qubit~\cite{blais:2004}.   This well defined environment makes
quantitative predictions for superconducting qubits more tractable.
This was shown in Ref.~\cite{wallraff:2005} where we have studied
Rabi oscillations in a superconducting qubit strongly coupled to a
superconducting transmission line resonator. Due to the detailed
understanding of the measurement process, we were able to make
quantitative predictions about the measured populations in the Rabi
oscillations and observe high visibility~\cite{wallraff:2005}
fringes. Moreover, as we showed in Ref.~\cite{schuster:2005},
populating the strongly coupled resonator with a coherent microwave
field can lead to a significant ac-Stark shift of the qubit, even in
the situation where detuning between the cavity and qubit
frequencies is large. Due to the shot noise in the number of photons
populating the resonator, this ac-Stark shift leads to
measurement-induced dephasing of the qubit. This is similar in
spirit to the experiment on Rydberg atoms in a 3D cavity reported in
Refs.~\cite{brune:1996, HarocheDis,
raimond:1997,MeuGleMaiAufNogBru05}.  In those time-domain
experiments, the visibility of Ramsey fringes was shown to decay
with an increase of the strength of dispersive coupling to the
cavity.

In this paper, we will expand on the theoretical model presented in
our experimental paper \cite{schuster:2005} (hereafter referred to
as the Letter) where we observed the AC stark shift and  measurement
induced dephasing in a circuit QED device. We will start in
Sec.~\ref{sec_circuit_qed_review} with a brief review of the
important features of circuit QED.  In Sec.~\ref{sec_exp} the
experimental results reported in the Letter will be reviewed.  We
then present in Sec.~\ref{sec_theory_main} two theoretical models
describing measurement-induced dephasing.  We first start with a
simple model which assumes Gaussian fluctuations of the qubit's
phase.  This is the model that was briefly presented in the Letter
to explain the experimental results.  We then present a more general
approach based on the positive P-representation~\cite{gardiner:2000}
which does not require the Gaussian assumption. For the experimental
parameters reported in the Letter, these two approaches give
identical results.  However, in the limit of strong coupling and
very high Q resonators, the second approach shows that qubit
spectrum will exhibit structure at several distinct frequencies due
to the underlying discrete energy levels of the cavity. That is, we
predict that the qubit spectrum will split into multiple peaks, with
each peak corresponding to a different number of photons in the
cavity. We will refer to this as {\em number splitting} of the qubit
spectrum. Experimental observation of this effect would be a direct
demonstration of number quantization in the dispersive regime.   We
also discuss how, by using irradiation which is off-resonant from
both the cavity and the qubit, one can obtain substantial ac-Stark
shifts without significant dephasing and how this could be used as
the basis of a phase gate for quantum computation.

\section{Cavity QED with superconducting circuits}
\label{sec_circuit_qed_review}

\subsection{Jaynes-Cummings interaction}

\begin{figure}[tp]
\centering \includegraphics[width=0.48\textwidth]{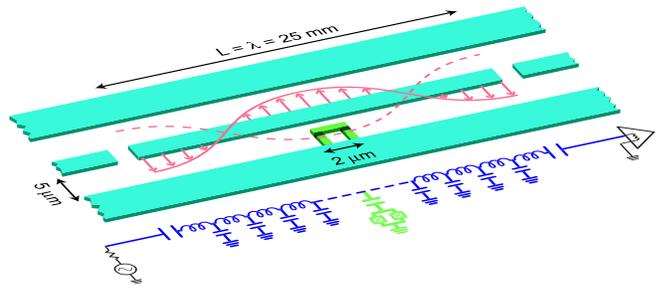}
\caption{(Color online) Schematic layout and lumped element version
of the  circuit QED implementation.  A superconducting charge qubit
(green) is fabricated inside a superconducting 1D transmission line
resonator (blue).} \label{fig_circuit}
\end{figure}

In this section, we briefly review the circuit QED architecture
first introduced in Ref.~\cite{blais:2004} and experimentally
studied in Refs.~\cite{wallraff:2004,schuster:2005,wallraff:2005}.
As shown in Fig.~\ref{fig_circuit}, the system consists of a
superconducting charge
qubit~\cite{bouchiat:98,makhlin:2001,devoret:2004} strongly coupled
to a transmission line resonator~\cite{frunzio:2004}.      Near its
resonance frequency $\omega_r$, the transmission line resonator can
be modeled as a simple harmonic oscillator composed of the parallel
combination of an inductor $L$ and a capacitor $C$.   Introducing
the annihilation (creation) operator $\aop^{(\dag)}$, the resonator
can be described by the Hamiltonian
\be
H_{r} = \hbar\wrr \ad \aop,
\ee
with $\wrr = 1/\sqrt{LC}$.  Using this simple model, one finds that
the voltage across the LC circuit (or, equivalently, on the center
conductor of the resonator) is $\vlc = V^0_\mathrm{rms}(\ad+ \aop)$,
where  $V^0_\mathrm{rms} = \sqrt{\hbar\wrr/2C}$ is the rms value of
the voltage in the ground state.  An important advantage of this
architecture is the extremely small separation $b \sim 5\:\mu$m
between the center conductor of the resonator and its ground planes.
This leads to a large rms value of the electric field
$E^0_\mathrm{rms} = V^0_\mathrm{rms}/b \sim 0.2$~V/m for typical
realizations~\cite{wallraff:2004,schuster:2005,wallraff:2005}.  As
illustrated in Fig.~\ref{fig_circuit}, by placing the qubit at an
antinode of the voltage, it will strongly interact with the
resonator through the large electric field $E^0_\mathrm{rms}$.

In the two-state approximation, the Hamiltonian of the qubit takes
the form
\be
H_q = -\frac{E_\mathrm{el}}{2}\sx - \frac{E_\mathrm{J}}{2}\sz,
\ee
where $E_\mathrm{el} = 4E_C(1-2n_g)$ is the electrostatic energy and
$E_\mathrm{J} = E_\mathrm{J,max} \cos(\pi \Phi/\Phi_0)$ the
Josephson energy.  Here, $E_C = e^2/2C_{\Sigma}$ is the charging
energy with $C_\Sigma$ the total box capacitance and $n_g =
C_gV_g/2e$ the dimensionless gate charge with $C_g$ the gate
capacitance and $V_g$ the gate voltage.  $E_ \mathrm{J,max}$ is the
maximum Josephson energy and $\Phi$ the externally applied flux,
with $\Phi_0$ the flux quantum.

Due to capacitive coupling with the center conductor, the gate
voltage $V_g = V_g^\mathrm{dc} + \vlc$ has a dc contribution
$V_g^\mathrm{dc}$ (coming from a dc bias applied to the input port
of the resonator) and a quantum part $\vlc$.  When working at the
charge degeneracy point $n_g^\mathrm{dc} = 1/2$ where dephasing is
minimized~\cite{vion:2002} and neglecting fast oscillating terms,
the resonator plus qubit Hamiltonian takes the Jaynes-Cummings
form~\cite{blais:2004}
\be
H_\mathrm{JC} = \hbar\wrr \ad \aop + \frac{\hbar\wa}{2}\sz - \hbar g
\left(\ad\smm+\spp \aop\right), \label{eq_Hjc}
\ee
where $\wa  = \ej/\hbar$ is the qubit transition frequency and $g =
e(C_g/C_\Sigma)V^0_\mathrm{rms}/\hbar$ is the coupling strength.

As shown in Ref.~\cite{blais:2004}, the qubit can be measured and
coherently controlled by applying microwaves, of frequency
$\omega_\rf$ and $\omega_\spec$ respectively, to the input port of
the resonator.  This can be described by the additional Hamiltonian
\be
H_\mathrm{D} = \sum_{j=\spec,\rf} \hbar\veps_{j}(t) \left(\ad e^{-i
\omega_jt}+ \aop e^{+i \omega_jt}\right), \label{eq_H_drive}
\ee
where $\veps_j(t)$ is the amplitude of the external drives at $\rf$
and spectroscopy frequencies.

\subsection{Dispersive regime}

In the situation where the qubit is strongly detuned from the
cavity, $|\Delta|  = |\wrr-\wa| \gg g$, the total Hamiltonian
$H_\mathrm{JC}+H_\mathrm{D}$ can be approximately diagonalized to
second order in $g/\Delta$ to yield the following quantized version
of the dynamical Stark shift Hamiltonian~\cite{blais:2004}
\be
\begin{split}
H_\mathrm{eff} = \: & \hbar\wrr \ad \aop
+ \frac{\hbar}{2}\left(\tilde \omega_a + 2 \chi \ad \aop \right)\sz\\
&
+ \sum_{j=\spec,\rf} \hbar\veps_{j}(t) \left(\ad e^{-i \omega_jt}+ \aop e^{+i \omega_jt}\right)\\
& +  \sum_{j=\spec,\rf} \frac{\hbar g\veps_j(t)}{\Delta} \left(\spp
e^{-i \omega_jt}+ \smm e^{+i \omega_jt}\right).
\end{split}
\label{eq_H_dispersive}
\ee
Here $\tilde\omega_a = \omega_a +\chi$ is the Lamb shifted qubit
frequency and we have defined $\chi = g^2/\Delta$.  The term
proportional to $\ad \aop\sz$ can be interpreted as a shift of the
qubit transition frequency depending on the photon number in the
resonator (ac-Stark shift) or as a pull  on the resonator frequency
by the qubit.  As will be shown later, quantum noise in the photon
number $\ad \aop$ leads to dephasing of the qubit.

Dephasing due to coupling to the cavity field was also studied using
Rydberg atoms coupled to a 3D  microwave
cavity~\cite{brune:1996,HarocheDis,raimond:1997}.   In this
experiment, atoms were sent one at a time through the  cavity and
interacted for a finite time with the field.  The state of the atoms
was finally read out by ionization~\cite{raimond:2001}.  Visibility
of the Ramsey fringes was measured as a function of the detuning
from the cavity, hence as a function of $\chi$.  Dephasing was shown
to increase with the strength of  dispersive coupling $\chi$ to the
cavity~\cite{HarocheDis}.   In this paper, we will instead consider
dephasing of the qubit due to the resonator field by looking at the
qubit spectrum as measured by transmission of the cavity field.

We note that, in practice, $\omega_\spec$ is chosen to be close to
$\tilde\omega_a$ and the last term of \eq{eq_H_dispersive} with
$\omega_\spec$ causes Rabi flopping of the qubit.  Moreover, as
further discussed below, we choose $\omega_\rf = \wrr - \Delta_r$ to
measure the state of the qubit, where $\Delta_r$ is the detuning of
the measurement probe from the bare cavity frequency.  In this
situation, the last term of \eq{eq_H_dispersive} with $\omega_\rf$
is largely detuned from the qubit and does not lead to qubit
transitions.  As first noted in the original proposal by Brune et
al.~\cite{brune:1990,brune:1992} this measurement Hamiltonian is
therefore highly quantum non-demolition~\cite{walls-milburn} with
respect to measurement of the qubit state.  Conversely, if the
dispersive coupling term is dominant in the Hamiltonian, then QND
measurement of photon number is also possible.  Physical
implementation of QND readout for superconducting qubits has been
achieved in the microwave
regime~\cite{wallraff:2004,Chiorescu:2004,schuster:2005,wallraff:2005},
but values of the coupling $\chi$ large enough to allow QND readout
in the dispersive regime for the photon number have not yet been
achieved in any system. However remarkable experiments on Rydberg
atoms have achieved photon number readout in a non-dispersive (i.e.
degenerate) regime~\cite{brune:1996,nogues:1999}.

\subsection{Damping}

Coupling to additional uncontrolled bath degrees of freedom leads to
energy relaxation and dephasing in the system.  Integrating out
these degrees of freedom leaves the qubit plus cavity system in a
mixed state $\rho(t)$ whose evolution can be described by the master
equation~\cite{walls-milburn}
\be
\begin{split}
\dot\rho  &=   \mathcal{L}\rho\\
& = -\frac{i}{\hbar}[H,\rho] +\kappa\mathcal{D}[\aop]\rho
+\gamma_1\mathcal{D}[\smm]\rho
+\frac{\gamma_\varphi}{2}\mathcal{D}[\sz]\rho,
\end{split}
\label{eq_master_eq}
\ee
where $\mathcal{D}[\hat L]\rho = \left(2 \hat L \rho \hat L^\dag
-\hat L^\dag \hat L \rho - \rho \hat  L^\dag \hat L\right)/2$
describes the effect of the baths on the system in the Markov
approximation. The last three terms of \eq{eq_master_eq} correspond
to loss of photons at rate $\kappa$, energy relaxation in the qubit
at rate $\gamma_1$ and pure dephasing of the qubit at rate
$\gamma_\phi$.

In the dispersive regime, the operators describing energy relaxation
and dephasing should be transformed in the same way as was done in
\erf{eq_H_dispersive}.  This leads to small corrections, of order
$(g/\Delta)^2$, to the master equation that are omitted here.

\section{Experimental results}
\label{sec_exp}

\begin{figure}[htb]
\centering
\includegraphics*[width=0.45\textwidth]{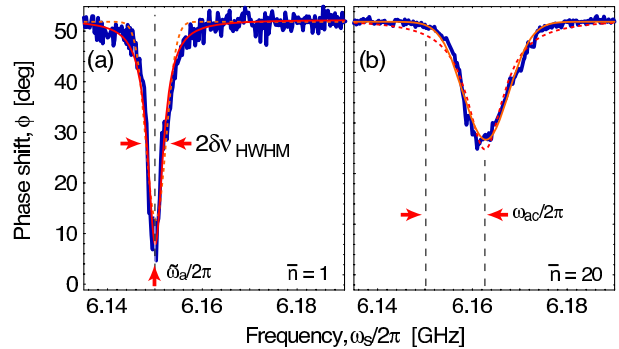}
\caption{(Color online) Measured spectroscopic lines (blue lines) at
(a) intra resonator photon number $\bar n \approx 1$ with fit to
lorentzian line shape (solid line) and at (b) $\bar n \approx 40$
with fit to gaussian line shape (solid line). Dashed lines are best
fits to (a) gaussian or (b) lorentzian line shapes, respectively.
The qubit transition frequency $\tilde\omega_a$ at low measurement
power, the half width half max $\delta\nu_{\rm{HWHM}}$ and the
ac-Stark shift $\omega_{\rm{ac}}$ of the lines are indicated.}
\label{fig_exp_lineshapes}
\end{figure}

In this section, we briefly review some of the experimental results
already presented in the Letter.  Only the results that are directly
discussed in the present paper will be presented and the details of
the experiment can be found in
Refs.~\cite{wallraff:2004,frunzio:2004,wallraff:2005,schuster:2005}.
In the Letter, we reported spectroscopic measurements of the qubit
as a function of measurement power.  The qubit spectroscopic line is
shown in Fig.~\ref{fig_exp_lineshapes} for two average photon
numbers $\bar n$ in the resonator, corresponding to two input
measurement powers.  The relevant experimental parameters are
$\Delta/2\pi = 105$~MHz, $g/2\pi= 5.8$~MHz, $\Delta_r/2\pi = 0$, and
$\kappa/2\pi = 0.57$~MHz and a dephasing time (in the absence of
power broadening) of $T_2 > 200$~ns.  These values correspond to a
relatively small cavity pull of $\chi/2\pi\approx 0.32$~MHz, or
$\chi/\kappa\approx 0.56$ in units of the cavity line width.  It is
important to note that at these relatively low detunings $\Delta$,
there can be a qubit contribution to the cavity line width.

\begin{figure}[tb]
\centering
\includegraphics[width=0.45\textwidth]{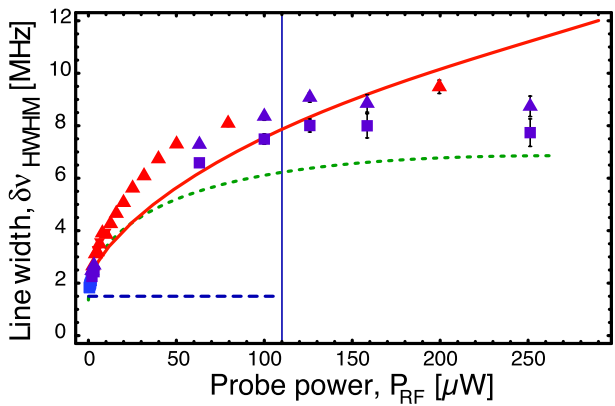}
\caption{(Color online)  Full, red curve: Measurement broadened
qubit line width $\delta\nu_{\rm{HWHM}}$ as a function of the input
measurement power or average photon number as predicted by the
lowest order dispersive approximation.
Green, dotted curve: Same as red but taking into account the
non-linear reduction in the cavity pull and plotted as a function of
input power.  The symbols are the experimental results. Symbols and
color scheme are described in the text.  The vertical line indicates
the critical photon number $\ncrit$. The parameters are those given
in section~\ref{sec_exp}. The blue dashed line is the calculated
HWHM for $\Delta_r/2\pi = 32$ MHz. It clearly shows that measurement
induced dephasing is small at large $\Delta_r$ where information
about the state of the qubit in the transmitted signal is also
small. This can thus be used as the basis of a phase gate. }
\label{fig_Deph_corrections}
\end{figure}

As discussed in the Letter, at low measurement power, the line shape
of the qubit spectrum is Lorentzian but as the measurement power
increases, the line shape approaches a Gaussian.  As shown in
Fig.~\ref{fig_Deph_corrections}, this is also seen in the dependence
of the half-width at half-maximum $\delta\nu_{\rm{HWHM}}$ of the
qubit line shape on $\bar n$ which goes from $\propto\bar n$ to
$\propto\sqrt{\bar n}$ as measurement power increases.  This figure
also shows theoretical results that will be discussed below.

In the Letter, we have already provided a theoretical explanation
for this behavior.  In section~\ref{sec_theory_main}, we review and
expand on this model.  We then explain how in the limit of very
large cavity pull, the results can be significantly different.

\section{ac-Stark shift}
\label{acStarktoymodel}

\begin{figure*}[tb]
\centering
\includegraphics[width=0.95\textwidth]{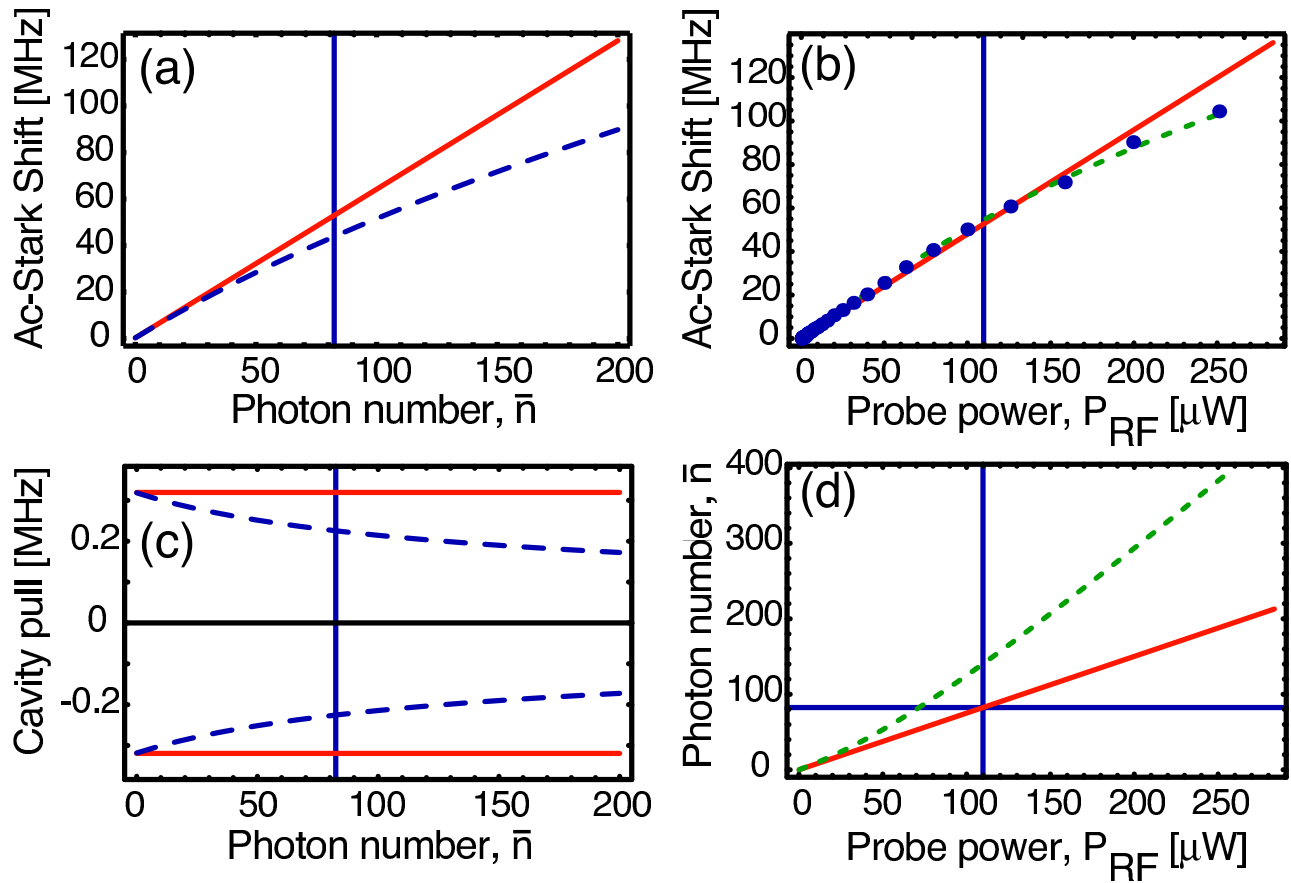}
\caption{(Color online)  (a) Full, red curve: ac-Stark shift as a
function of the average intra-cavity photon number using the lowest
order dispersive approximation.  Dashed, blue curve: ac-Stark shift
as a function of photon number calculated from the exact eigenvalues
of the Jaynes-Cummings Hamiltonian.
 (b) Blue dots: experimentally
measured ac-Stark shift as a function of external microwave input
power. Full, red curve:  Predicted ac-Stark shift within the lowest
order dispersive approximation.  The conversion factor from photon
number to external microwave drive power was determined by fitting
the red curve to the linear portion of the data at low power. Green,
dotted curve: Same as red but taking into account the non-linear
reduction in the cavity pull (see part c) and the non-linear
increase in the average photon number (see part d) with microwave
drive power. For the particular experimental parameters these two
effects almost cancel each other out and result in the green dotted
line being nearly linear out to much greater input powers than
expected.
 (c) Cavity pull as a function of average photon number $\bar n$.
The red solid line is the result of the dispersive approximation
($\pm\chi$) while the dashed blue curve is obtained from the exact
eigenvalues of the Jaynes-Cummings model.
(d) Average photon number as a function of input power.  The full
red line is the result of the lowest order dispersive dispersive
approximation~\eq{eq_ave_photon_power} fit to the data in (b) at low
power. The dotted green line is the non-linear model with $\chi$
replaced by $\chi(\bar{n})$ in~\eq{eq_ave_photon_power}  The
vertical line in all plots indicates the critical photon number
$\ncrit=\Delta^2/4g^2$ which indicates the scale at which the lowest
order dispersive approximation breaks down.
  For the experimental parameters
 (given in section~\ref{sec_exp}) $\ncrit\sim 82$ which corresponds
 (within the lowest order dispersive approximation) to $P_{\rm RF}\sim 110\mu$W.}
\label{fig_Stark_corrections}
\end{figure*}

In the lowest order dispersive approximation  [\eq{eq_H_dispersive}]
the predicted ac-Stark shift $\omega_{\rm ac} = 2 \chi \bar{n}$
 will be a linear function of the mean photon number, $\bar{n}$.
However, this approximation only holds at low photon numbers and
breaks down on a scale given by the critical photon number
$\ncrit=\Delta^2/4g^2$~\cite{blais:2004}.   This is illustrated in
Fig.~\ref{fig_Stark_corrections}a) where the ac-Stark shift,
calculated from the lowest order dispersive approximation (red solid
line) and the exact eigenvalues (blue dashed line) of the
Jaynes-Cummings model~\cite{blais:2004} are plotted for the
experimental parameters. Also shown in this figure is $\ncrit$
(vertical blue line)  which for the experimental parameters is about
82 photons. Here we see that as $\bar{n}$ increases, the exact Stark
shift begins to fall below the lowest order dispersive approximation
even before $\bar{n}$ reaches $\ncrit$.

In Fig.~(\ref{fig_Stark_corrections}b) the experimental results
(solid blue points) are plotted as a function of probe power,
$P_{\rm RF}$ (extending up to powers larger than those presented in
the Letter).
%
To convert between $\bar n$ and $P_{\rm RF}$, we assume that $P_{\rm
RF} =\lambda \hbar\omega_{\rm RF} p$ where $p$ is the photon flux at
the resonator and $\lambda$ is a scaling factor that takes into
account the large attenuation that is placed between the probe
generator and the resonator (to eliminate black body radiation).
\null From the lowest order dispersive approximation, the average
photon number when driving the cavity at $\Delta_r = 0$ is
\be
\bar n = \frac{p\kappa/2}{(\kappa/2)^2+\chi^2}.
\label{eq_ave_photon_power}
\ee
By using the lowest order dispersive approximation for the ac-Stark
shift and the line of best fit to the experimental points (in the
linear regime at low power) $\lambda$ can be determined. Doing this
gives the red solid line in Figs.~\ref{fig_Stark_corrections}b) and
d). The calibration shows that $\ncrit$ occurs at $\approx 110\mu$W.
The experimental results clearly show the breakdown of the lowest
order dispersive approximation.   The data points fall below the
linear prediction but not nearly as much as the blue dashed curve in
(a) predicts.  In fact, the data points follow fairly closely the
linear in $\bar n$ dependence of the lowest order dispersive
approximation in \eq{eq_H_dispersive} for larger powers than
expected (up to and well above $\ncrit$).

 It is possible to understand why the experiment agrees with the
 simple dispersive approximation for larger probe powers than expected
 by considering the following simple model.  We assume that, at
 these large powers, the ac-Stark shift is still given by the
 dispersive approximation $2\chi\bar n$, but we now take into
 account the non-linear cavity pull (which is $\chi$ at low $\bar
 n$). From the eigenvalues of the Jaynes-Cummings Hamiltonian the
 cavity pull can be calculated as a function of  $\bar n$. This is
 shown in  Fig.~\ref{fig_Stark_corrections}c).  From this figure, we
 see that the cavity pull reduces as the number of photons in the
 resonator is increased.  We thus replace $\chi$ by $\chi(\bar n)$.
 The second aspect of our simple model is the non-linear dependence
 of the average photon number $\bar n$ with input power $P$ due to
 the power-dependence of the cavity frequency.
 To account for this we simply replace $\chi$ in
 \erf{eq_ave_photon_power} with $\chi(\bar{n})$ and $\bar n$ becomes
 a non-linear function of input power. This non-linear dependence of
 the photon number on the input power is illustrated in
 Fig.~\ref{fig_Stark_corrections}d) as the green dotted line.  This
 is a precursor to bistability in this system~\cite{walls-milburn}.
 Using these two expressions, we have for our simple model of the
 non-linear ac-Stark shift $2\chi(P)\bar n(P)$.  This expression is
 plotted in Fig.~\ref{fig_Stark_corrections}b) (green dotted line)
 with a new scaling factor $\lambda'\approx 0.905 \lambda$
 calculated by the best fit for
 the experimental data (here we use the complete data set). This
 simple model produces a result that is linear for a larger range of
 powers and is closely consistent  with the experimental results. It happens that
  for the particular experimental parameters, the two non-linear effects almost cancel
 each other out and result in the green dotted line being more
 linear than expected.

 We emphasize that in the Letter, only the
 low power (below $\ncrit$) part of the ac-Stark shift was studied
 and was fit only with the linear dispersive model.  Comparison with
 the results of the
 non-linear model shown here in Fig.~\ref{fig_Stark_corrections}d)
 shows that the calibration of the cavity
 photon number in terms of the drive power
 is low by approximately 50\% at the highest power shown in Fig.~5 of the
 Letter.  We also emphasize that our treatment here of the
 non-linearities is only approximate.


\section{Measurement-induced dephasing}
\label{sec_theory_main}

By monitoring the transmission of the cavity using heterodyne
detection, one has access to the average of the cavity field
$\langle \aop(t)\rangle$.   As shown in Ref.~\cite{blais:2004}, the
phase $\phi(t) = \arg\{\langle \aop(t)\rangle\}$ is directly related
to the population of the qubit $\langle \sz(t)\rangle$.  As a
result, by recording the phase $\phi(t)$ as a function of the
excitation frequency $\omega_\spec$, one has access to the
absorption spectrum of the qubit~\cite{gardiner:2000}
\be
S(\omega) = \frac{1}{2\pi}\int_{-\infty}^\infty dt~ e^{i\omega
t}\langle\smm(t)\spp(0)\rangle_s,
\ee
where the subscript $s$ implies that the expectation value is taken
in the steady state.  The dephasing rate can be determined through
the half width at half maximum of $S(\omega)$~\cite{abragam}.

Using the quantum regression formula~\cite{gardiner:2000}, the
correlation function $\langle\smm(t)\spp(0)\rangle_s$ can be
evaluated as
\be\label{corr}
\langle\smm(t)\spp(0)\rangle_s = \left\{
\begin{aligned}
&\mathrm{Tr}[\smm e^{\mathcal{L}t}\spp\rho_s]  \quad &t > 0\\
&\mathrm{Tr}[\spp e^{-\mathcal{L}t}\rho_s\smm] \quad &t < 0,
\end{aligned}
\right.
\ee
where $\rho_s$ is the steady-state density matrix. This allows us to
rewrite the spectrum as
\begin{eqnarray}
S(\omega) = \frac{1}{\pi}{\rm Re}\Big{[}\int_{0}^\infty dt~
e^{i\omega t}\langle\smm(t)\spp(0)\rangle_s\Big{]}.
\end{eqnarray}

We start by calculating this spectrum by assuming gaussian
statistics for the qubit's phase noise.  This approximation is
sufficient for the experimental parameters considered above but
breaks down in the situation where the cavity pull is large $\chi
\gg \kappa$.  Moreover, this simple approach has the advantage of
presenting the essential physics in a transparent way.  We then show
how to go beyond the gaussian approximation by using the positive-P
function approach~\cite{gardiner:2000}.

\subsection{Gaussian approximation for the phase}
\label{sec_Gaussian}

As mentioned above, quantum fluctuations $\delta n$ in the photon
number around its average value $\bar n$ will lead to accumulation
of a random relative phase between the amplitudes of the two basis
states of the qubit and hence to dephasing.   We first consider the
situation where the qubit is prepared in a superposition
$\ket{\psi(0)} = \left(\ket{g}+\ket{e}\right)/\sqrt{2}$ of its basis
states and that the cavity is populated by a coherent state with
average photon number $\bar n$.  As it evolves under the dispersive
Hamiltonian \eq{eq_H_dispersive}, the qubit superposition picks up a
relative phase factor
\be
\varphi(t) = \tilde\omega_a  t + 2\chi \int_0^t dt' n(t'),
\ee
where $\tilde\omega_a = \omega_a + \chi$ is the Lamb shifted qubit
transition frequency.  It is convenient to express the second term
as its mean value plus fluctuations about the mean
\begin{equation}
\varphi(t) \equiv \bar \varphi + \delta\varphi(t) =  \tilde\omega_a
t + 2\chi \bar n t + 2\chi \int_0^t dt' \delta n(t'),
\end{equation}
with $\bar n$ the average photon number in the cavity leading to the
ac-Stark shift (see Fig.~\ref{fig_Stark_corrections}) and $\delta
n(t)$ the random excursions about this mean.

In a frame rotating at the Lamb and ac-Stark shifted qubit
transition frequency, we obtain for the correlation function ($t>0$)
\be
\begin{split}
\langle\smm(t)\spp(0)\rangle_{s}
& = \mathrm{Tr}[\smm e^{\mathcal{L}t}\spp\rho_{s}]\\
& = \mathrm{Tr}[\smm e^{\mathcal{L}t} (\ket{e}\bra{g})]\\
& = e^{-\gamma_2 t}\langle e^{-i\delta\varphi(t)}\rangle,
\end{split}
\ee
where $\gamma_2 = \gamma_1/2 + \gamma_\varphi$. This is the
off-diagonal component of the reduced qubit density matrix. Assuming
gaussian statistics for the phase $\delta\varphi(t)$, the cumulant
expansion is exact and we obtain \cite{Gar85}
\begin{equation}
\begin{split}
&\langle\smm(t)\spp(0)\rangle_{s} \\
&\approx e^{-\gamma_2 t}e^{-\frac{1}{2}\langle \delta\varphi^2 \rangle} \\
&= e^{-\gamma_2 t}\exp\left[ -2\chi^2 \iint_0^t dt_1 dt_2 \langle
\delta n(t_1)\delta n(t_2)\rangle \right]. \label{eq_exp_ave}
\end{split}
\end{equation}

This expression involves the photon-photon time correlator which for
a two-sided symmetrically damped driven cavity takes the
form~\cite{blais:2004}
\begin{equation}
\langle \delta n(t_1)\delta n(t_2)\rangle = \bar n
e^{-\frac{\kappa}{2}|t_1-t_2|}, \label{eq_photon_corr}
\end{equation}
leading to
\begin{equation}
\begin{split}
&\langle\smm(t)\spp(0)\rangle_s\\
&  =e^{-\gamma_2 t} \exp\left[ -4\bar n\theta_0^2 \left\{
\frac{\kappa |t|}{2}-1 +\exp\left(-\frac{\kappa |t|}{2}\right)
\right\} \right],
\end{split}
\label{eq_phase_decay}
\end{equation}
where $\theta_0 = \tan^{-1} 2\chi/\kappa \approx 2\chi/\kappa$ is
the magnitude of the accumulated phase shift for the transmitted
photons due to the coupling with the qubit in the small pull
approximation ($\chi \ll \kappa)$ and at $\omega_\rf=\wrr$.

We now consider two simple limits of the above result.  First, in
the situation where the mean cavity photon number $\bar n$ is small,
fluctuations of the photon number will only weakly contribute to
dephasing.  In this situation phase decay occurs on a long time
scale with respect to $1/\kappa$.  In this limit, $\exp(-\kappa
|t|/2) \approx 0$ and \eq{eq_phase_decay} reduces to
\begin{equation}
\langle\smm(t)\spp(0)\rangle_s \approx \exp\left[-\left\{ \gamma_2 +
2\bar  n\kappa\theta_0^2 \right\} |t|\right].
\label{eq_lorentz_time}
\end{equation}
In this situation, the measurement induced-dephasing only adds to
the intrinsic dephasing rate $\gamma_2$.  This is because, in this
long time limit, the phase undergoes a random walk process leading
to an exponential decay of the coherence.  The Fourier transform of
this expression leads to a Lorentzian spectrum with half-width at
half maximum $\gamma_2+\tGm$, where $\tGm = 2\kappa\bar n\theta_0^2$
is the measurement induced dephasing rate in the small pull limit
(see next section).

On the other hand, in the large $\bar n$ limit, phase decay can
occur on a time scale much shorter than the cavity lifetime, $t \ll
\kappa^{-1}$.   Indeed, since $\bar n$ is large, a small fraction of
$\bar n$ leaking out of the cavity in a time $t \ll \kappa^{-1}$
conveys enough information to infer the state of the qubit and hence
to dephase it completely~\cite{clerk:2003}.  In this situation,
expanding $\exp(-\kappa |t|/2)$ in \eq{eq_phase_decay}, we obtain
\begin{equation}
\langle\smm(t)\spp(0)\rangle_s \approx \exp\left[- \gamma_2 |t| -
2\bar n \chi^2 t^2 \right]. \label{eq_gauss_time}
\end{equation}
The large $\bar n$ limit does not lead to an exponential decay and
the spectrum will be a convolution of a Lorentzian and a Gaussian.
This corresponds to inhomogeneous broadening of the qubit due to the
Poisson statistics of the coherent state populating the cavity.  In
this situation, the half width at half maximum therefore scales as
$\sqrt{\bar n}$.

Using the full expression \eq{eq_phase_decay} and moving back to the
lab frame, we obtain for the spectrum of the qubit in the Gaussian
approximation for the phase
\begin{equation}
\begin{split}
\tilde S(\omega)
= & \frac{1}{2\pi}\sum_j \frac{(-\frac{2\tGm}{\kappa})^j}{j!}
\frac{\frac{1}{2}\tilde\Gamma_j}{(\omega-\tilde\omega_a-2 \bar n
\chi)^2+\left(\frac{1}{2}\tilde\Gamma_j\right)^2},
\end{split}
\label{eq_lineshape_gaussian_approx}
\end{equation}
where $\tilde\Gamma_j=2(\gamma_2+\tGm)+j\kappa$.    The
spectroscopic line shape is given by a sum of Lorentzians, all
centered on the ac-Stark shifted qubit transition but of different
widths and weights.

As expected from the above discussion,  we see from
\eq{eq_lineshape_gaussian_approx} that if the measurement rate
$\tGm$ is much smaller than the cavity decay rate $\kappa/2$, then
only a few terms in the sum contribute and the spectrum is
Lorentzian.  On the other hand, when the measurement rate is fast
compared to the cavity damping, the spectrum will be a sum of many
Lorentzians, resulting in a gaussian profile.  In this situation,
dephasing occurs before the cavity has had time to significantly
change its state, leading to inhomogeneous broadening as discussed
above.

The expression \eq{eq_lineshape_gaussian_approx} for the spectrum
can be summed analytically but yields an unsightly result which is
not reproduced here.  To compare with the experimental results, we
evaluate numerically the half-width at half maximum from
$S(\omega)$.  The results are plotted as a function of probe power
in Fig.~\ref{fig_Deph_corrections} (full red line).    The agreement
with the experimental results (symbols) is good, especially given
that there are no adjustable parameters apart from $\gamma_2$ which
only sets the value of the dephasing at $\bar n =0$.    In this
figure we have included more experimental points, for higher powers,
than presented in Fig. 5 of the Letter. The experimental points
presented here are obtained by fitting a Lorentzian (blue squares)
and a Gaussian (red triangles) to the experimentally measured
spectroscopic line (see Fig.~\ref{fig_exp_lineshapes}).  We then
keep the fit which has the smallest variance or both points if the
variances are approximately the same (purple squares and triangles).
The error bars are the standard errors on the half-width half-max
obtained from the fit.  This approach to extracting the error bar is
different from what was presented in Fig. 5 of the Letter.  In that
case, the error bars represented the systematic
difference between the Lorentzian and Gaussian fits and the points
were the average value. From Fig.~\ref{fig_Deph_corrections} (this
paper) we see that the first 7 points fit best to a Lorentzian while
the later points fit best to a Gaussian, except for the higher
powers where the error in the fit is approximately the same. The
predicted crossover from Lorentzian to Gaussian is clearly seen.

There are several potential sources of discrepancy between the
experimental and the theoretical results.  One of them is the
breakdown of the lowest order dispersive approximation.  Using the
same simple non-linear model as in section~\ref{acStarktoymodel}, we
plot in Fig.~\ref{fig_Deph_corrections} the half-width at half
maximum as a function of input power (green dashed line).  The
effect of this correction is to reduce the width of the spectroscopy
peaks.

The breakdown of the dispersive approximation can be seen by the
dispersive result (red full line) over-estimating the width at high
powers.  However, while the simple non-linear model used here does
correctly show saturation of the width at high powers, it is not a
full treatment of the Jaynes-Cummings Hamiltonian and should not be
considered too seriously.  A complete investigation of the behavior
of the system at very large photon numbers will require numerical
investigation which is beyond the scope of this paper.

A further possible source of discrepancy comes from the fact that a
constant spectroscopy power was used.  Since the effective coupling
strength $\chi$ is not constant with probe power, the effect of the
spectroscopy power on the qubit will change with measurement probe
power. As a result, spectroscopic power
broadening~\cite{schuster:2005} will also depend on measurement
probe power. This effect has been taken into account in the green
dashed curve of Fig.~\ref{fig_Deph_corrections} but only to the
accuracy of our simple model.  Finally, environmental noise due to
two-level systems activated at large photon number could be an
additional cause of discrepancy.

\subsection{Beyond the gaussian approximation}

To go beyond the gaussian approximation made in the last section, we
solve the master equation \eq{eq_master_eq} using the positive
$P$-function method~\cite{walls-milburn}.  Following
Ref.~\cite{Milburn-Wiseman}, we first write the qubit-cavity density
matrix as
\begin{equation}\label{EQ.StateMatrix}
\rho=\hat\rho_{ee}\ket{e}\bra{e}+\hat\rho_{gg}\ket{g}\bra{g}+\hat\rho_{eg}\ket{e}\bra{g}+\hat\rho_{ge}\ket{g}\bra{e},
\end{equation}
where $\hat\rho_{ij}$ acts only in the cavity Hilbert space.  As
shown in appendix~\ref{sec_appendix_P}, this leads to four coupled
differential equations for the operators $\hat\rho_{ij}$.  In the
absence of qubit mixing ($T_1$ processes), solving these coupled
equations yields the time-evolved full density matrix
\begin{eqnarray}\label{EQ.Sol}
 \rho(t) &=&
 c_{ee}(0)\ket{e}\bra{e}\otimes\ket{\alpha_+(t)}\bra{\alpha_+(t)}\nonumber\\
 &+&
 c_{gg}(0)\ket{g}\bra{g}\otimes\ket{\alpha_-(t)}\bra{\alpha_-(t)}\nonumber\\
 &+&
 c_{eg}(t)\ket{e}\bra{g}\otimes\ket{\alpha_+(t)}\bra{\alpha_-(t)}\nonumber\\
 &+&
 c_{ge}(t)\ket{g}\bra{e}\otimes\ket{\alpha_-(t)}\bra{\alpha_+(t)},
\end{eqnarray}
where
\begin{equation}\label{EQ.c}
c_{eg}(t)=\frac{a_{eg}(t)}{\bra{\alpha_-(t)}\alpha_+(t)\rangle}
\end{equation}
describes the decay of the qubit phase coherence.  In the above
expression, we have
\begin{equation}\label{EQ.a}
  {a}_{eg}(t)
  = {a}_{eg}(0)
  e^{-i(\tilde\omega_a-i\gamma_2) t}
  e^{-i 2 \chi \int_0^t  \alpha_+(t')\alpha_-^*(t')dt'}
\end{equation}
and ${a}_{ge}(t) = {a}_{eg}^*(t)$ with
\begin{equation}
    \alpha_+(t)= \alpha_+^{\rm s} +\exp[-(\kappa/2+i\chi+i\Delta_r)t][\alpha_+(0)-\alpha_+^{\rm
    s}],
\end{equation}
where $\alpha_+^{\rm s}=-i \veps_\rf /(\kappa/2 +i \chi+i\Delta_r)$.
Moreover, we have
\begin{equation}
   \alpha_-(t)= \alpha_-^{\rm s} +\exp[-(\kappa/2-i\chi+i\Delta_r)t][\alpha_-(0)-\alpha_-^{\rm
    s}],
\label{eq_alpha_minus}
\end{equation}
with $\alpha_-^{\rm s}=-i \veps_\rf /(\kappa/2 -i \chi+i\Delta_r)$.
In these expressions, $\alpha_\pm^{s}$ represent the steady-state
value of the field $\langle\hat a\rangle$ given that the qubit is
either in its ground ($-$) or excited ($+$) state.  Recall that
$\Delta_r$ is the detuning of the measurement beam from the bare
cavity frequency.

\begin{figure}[tb]
\begin{center}
\includegraphics[width=0.45\textwidth]{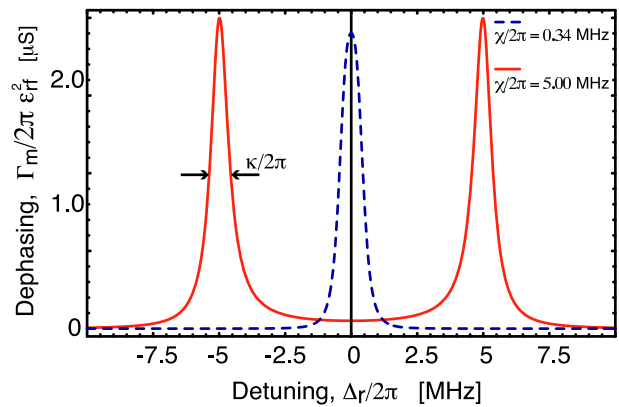}
\caption{\label{fig_dephasing} (Color online) Measurement-induced
dephasing rate $\Gm$ as a function of the detuning $\Delta_r$
between the bare resonator frequency $\omega_r$ and the measurement
drive frequency $\omega_d$.  The dephasing rate is divided by the
measurement power $\veps_\rf^2$  since it only changes the overall
scale and not the structure of $\Gm$.  The blue (dashed) line
corresponds to the experimental parameters given in
section~\ref{sec_exp}.  The red (full) line has the same parameters
but a larger cavity pull $\chi/2\pi=5$ MHz.}
\end{center}
\end{figure}

In the limit  $\kappa t\gg 1$ discussed previously,  the decay of
${a}_{eg}(t)$  is given by
\begin{eqnarray}
  {a}_{eg}(t) &\sim& a_{eg}(0)\exp[-(\gamma_2 + \Gm) t],
\end{eqnarray}
where $\Gm$
\be
\label{Gamma} \Gm = -2 \chi \mathrm{Im}
[\alpha_+^\mathrm{s}{\alpha_-^\mathrm{s}}^*]
=\frac{(\bar{n}_++\bar{n}_-)\kappa\chi^2}{{\kappa^2/4+\chi^2+\Delta_r^2}}
\ee
is the generalized measurement-induced dephasing rate.  In this
expression,
\begin{eqnarray}
    \bar{n}_\pm
    &=&|\alpha_\pm^{\rm s}|^2
    =\frac{\varepsilon_{\rm rf}^2}{\kappa^2/4+(\Delta_r\pm\chi)^2}
\end{eqnarray}
is the stationary average number of photons in the cavity when the qubit
is in the excited ($+$) or ground ($-$)  state.  At small pulls, as is
the case for the experimental parameters quoted in section~\ref{sec_exp},
the measurement-induced dephasing rate is largest at
$\Delta_r=0$ (see Fig.~\ref{fig_dephasing}, blue dashed line).
At large pulls, cavity transmission decreases at $\Delta_r=0$ and,
as illustrated in Fig.~\ref{fig_dephasing} (red solid line), the maximum
dephasing rate then occurs at $\Delta_r=\pm\sqrt{\chi^2-\kappa^2/4}$.
As we increase $\chi$, the information about the state of the qubit is
conveyed more by the amplitude than the phase of the transmitted beam.

\null From the above, we see explicitly that by introducing a probe
($\varepsilon_{\rm rf}$) we cause the coherence terms $a_{eg}(t)$ to
exponentially decay at a rate $\Gamma_m$ thereby leaving the system
in a mixed state with perfect correlation between the eigenstates of
$\hat\sigma_z$ and the pointer states $\ket{\alpha_+}$ and
$\ket{\alpha_-}$.  As a result, if the pointer states are well
separated in phase space,  we can regard the cavity as a meter which
performs a von Neumann projective measurement  of the qubit
observable $\hat{\sigma}_z$.

A measure of the distinguishability of the cavity states is
$D=|\alpha_+-\alpha_-|^2$ \cite{walls-milburn}. If this is large
such that $|\bra{\alpha_-} \alpha_+\rangle| = \exp [- D]$ is small,
then the two pointer states are well separated and easily
distinguishable. In the steady-state, $D$ is given by
\begin{eqnarray}
D_{\rm s} = \frac{
2(\bar{n}_++\bar{n}_-)\chi^2}{{\kappa^2/4+\chi^2+\Delta_r^2}},
\end{eqnarray}
which is related to $\Gm$ in the following way
\begin{eqnarray}
\Gm=\frac{D_{\rm s}\kappa}{2}. \label{eq_Gm}
\end{eqnarray}
That is, as the measurement becomes more projective in the
$\hat\sigma_z$ basis, the qubit dephases faster.  This is a clear
example of measurement induced dephasing and of the fundamental
limit which exists between acquiring information about a quantum
system and dephasing of that system~\cite{clerk:2003}.    Note that,
as shown appendix~\ref{sec_appendix_meas},  $D_{\rm s}$ can also be
related to the measurement time.  The present system is a factor of
4 away from the quantum limit.  One factor of 2 comes from the fact
that we are using a symmetric resonator and only looking at the
transmission.  Half of the information is lost in the unmeasured
reflected signal~\cite{blais:2004}.  The other factor of two comes
from the use of heterodyne rather than homodyne detection.
As a result and as shown in this appendix, the quantum limit can be
reached by using asymmetric resonators and homodyne detection of the
transmitted field.

\subsubsection{The qubit absorption spectrum}

To evaluate the qubit's spectrum, we first need to calculate the
correlation function $\langle\smm(t)\spp(0)\rangle_s$.
%
Note that (as discussed in appendix \ref{sec_appendix_P})
in calculating this particular correlation function we do not need to assume that $T_1$ is
infinite. This is because it only depends on the off-diagonal coherences and these are not mixed by a $T_1$ process. Using \erf{corr} and the
above results, the correlator
 can be shown
to be~($t>0$)
\begin{equation}
\langle\smm(t)\spp(0)\rangle_s=  a_{eg}(t),
\end{equation}
with the initial condition $\alpha_+(0)=\alpha_-(0)=\alpha_-^{\rm
s}$. Using \erf{EQ.a} with the above initial condition yields
\begin{eqnarray}\label{EQ.Correlation2}
\langle\smm(t)\spp(0)\rangle_s &=& \exp[-(\gamma_2 +\Gm)t
-i(\tilde\omega_a+B)t]\nonumber\\ &\times& \exp[-A
e^{-(\kappa/2+i\chi+i\Delta_r)t}]\exp[A] ,\nl
\end{eqnarray}
where
\begin{eqnarray}
A &=& -i 2\chi \frac{(\alpha_-^{\rm s}-\alpha_+^{\rm
s}){\alpha_-^{\rm s}}^*}
{\kappa/2+i\chi +i \Delta_r}\nn\\
&=& D_{\rm s}\frac{\kappa/2-i\chi- i \Delta_r}{\kappa/2+i\chi + i \Delta_r},\\
B &=& 2\chi{\rm Re}[\alpha_+^{\rm s}{\alpha_-^{\rm s}}^*]=
\chi(\bar{n}_++\bar{n}_-)- \chi D_{\rm s}. \label{eq_B}
\end{eqnarray}
\null From the above expression, we see that the time dependence of
the correlation function is given by three terms.  The first one
involves $\Gm$ and is simply the Lorentzian part of the
measurement-induced dephasing spectrum we have seen before.  The
second is a frequency shift $B$ which contains a negative term
$-\chi D_{\rm s}$ which gives rise to negative frequency
contributions in the spectrum.  The last relevant term goes as  $A
\exp[\{\kappa/2+i(\chi+\Delta_r)\}t]$ and gives rise to
non-Lorentizian spectra.

\null From the expression for the correlation function, it is simple
to obtain the spectrum:
\begin{equation}
S(\omega) = \frac{1}{\pi}
\sum_{j=0}^\infty\frac{1}{j!}\mathrm{Re}\left[ \frac{(-A)^j
e^A}{\Gamma_j/2-i(\omega-\omega_j)}\right],
\label{eq_lineshape_non_gaussian_approx}
\end{equation}
where $\Gamma_j = 2(\gamma_2+\Gamma_m)+j\kappa$ and $\omega_j
=\tilde\omega_a+B + j (\chi+\Delta_r)$.  The spectrum, as in the
Gaussian approximation, can be written as a sum over different
photon numbers $j$. In the limit of $(\Delta_r +\chi)$ much
different from $\kappa/2$, the spectrum is a sum of Lorentzians with
decay rate $\Gamma_j/2$.   However unlike
\eq{eq_lineshape_gaussian_approx} where each Lorentzian is centered
at the ac-Stark shifted frequency, here each peak has its own
frequency shift $\omega_j$.  As a result, the full theory predicts
that the spectrum need not be symmetric whereas in the  Gaussian
theory only symmetric spectra are possible. Furthermore, in the
limit that $\chi$ is much larger than $\kappa$, $A\rightarrow D_{s}$
and the spectral weights become Poisson distributed with mean $D_s$.
The peaks are separated by $(\chi+\Delta_r)$ and the first one is at
the frequency $\tilde\omega_a + B$. That is, the average frequency,
which is the ac-Stark shift, occurs at
\begin{equation}
\omega_{\rm ac} = B+(\chi+\Delta_r)D_s = 2\chi n_-.
\end{equation}
Taking this limit further with $\chi$ much larger than the widths
$\Gamma_j$, the individual peaks will become distinguishable.   This
is discussed further below.

It is interesting to point out that in the limit of large
$\bar{n}_{\rm -}$ (or $\bar{n}_{\rm +}$) the results obtained here
and those obtained in the Gaussian approximation agree.  This can be
seen by expanding the exponent $A
\exp[\{\kappa/2+i(\chi+\Delta_r)\}t]$ in \eq{EQ.Correlation2} to
order $t^2$:
\be
\begin{split}
\label{EQ.Correlation3} \langle\smm(t)\spp(0)\rangle_s &\approx
\exp\Big[-(\gamma_2+i\tilde\omega_a+i2\chi\bar{n}_-)t \\
&-\frac{D_{\rm s}}{2}(\kappa^2/4+(\Delta_r+\chi)^2)|t|^2\Big].
\end{split}
\ee
For $\Delta_r = 0$, $\bar{n}_- = \bar{n}_+ \equiv \bar n$ and we
recover \erf{eq_gauss_time} in the lab frame.  As a result, in the
large $\bar{n}$ limit, both the gaussian approximation and the above
theory converge to give the same gaussian spectrum. It is only when
$\bar{n}$ is small that the theories have different predictions.
This will be discussed further in section~\ref{numSec}.

The half-width at half maximum of the spectrum obtained from the
full expression~\eq{eq_lineshape_non_gaussian_approx} is plotted at
$\Delta_r = 0$ as a function of $\bar n$ in
Fig.~\ref{fig_Deph_corrections} using the experimental parameters
given in section~\ref{sec_exp}.  Since the experiment was done in
the limit $\chi < \kappa/2$, the results obtained
from~\eq{eq_lineshape_non_gaussian_approx} cannot be distinguished
from those obtained from the Gaussian
approximation~\eq{eq_lineshape_gaussian_approx}. This is because in
the small cavity pull limit  $\Gamma_m \rightarrow \tilde\Gamma_m $
and therefore $A \rightarrow 2 \tilde\Gamma_m/\kappa  $,  $B
\rightarrow 2\bar{n} \chi$.  That is, to see the break down of the
gaussian approximation at small $\bar{n}$ requires a larger cavity
pull.


\subsubsection{Phase-gate}

As discussed above and in appendix~\ref{sec_appendix_meas},
measurement causes dephasing of the qubit.  This is a clear
illustration of the Heisenberg type relation between rate of
information gain and dephasing $\Gm$~\cite{clerk:2003}.  However,
irradiation at the $\rf$ frequency does not have to induce dephasing
of the qubit.  Indeed, the qubit pulls the resonator frequency up or
down causing a state dependent phase shift for photons near the
cavity frequency.  But, in the low $\chi$ limit, photons off
resonant from the resonator have phase shifts nearly independent of
the qubit state.  These photons do not become entangled with the
qubit, and hence do not cause dephasing.

\begin{figure}[tb]
\begin{center}
\includegraphics[width=0.45\textwidth]{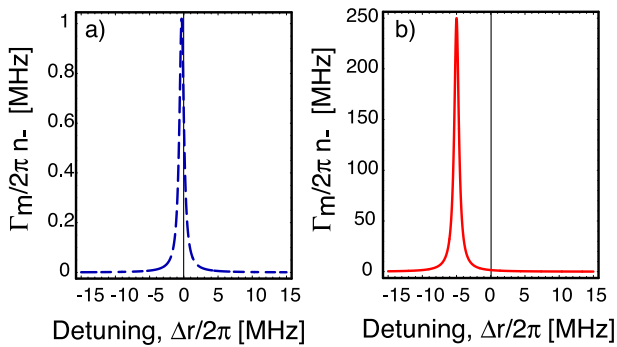}
\caption{\label{fig_dephasing2} (Color online) Dephasing rate $\Gm$
as a function of the detuning $\Delta_r$  between the bare resonator
frequency $\omega_r$ and the measurement drive frequency $\omega_d$
for fixed $\bar{n}_-=2$.  a) Experimental parameters given in
section~\ref{sec_exp}.  b) Same as a) but with a larger cavity pull
$\chi/2\pi=5$ MHz.}
\end{center}
\end{figure}

This can be understood more quantitatively in
Fig.~\ref{fig_dephasing}, where it can be seen that the dephasing
rate is significant only on a frequency range $\kappa$ around the
pulled resonator frequency.  In this Figure, we have fixed the input
power and scanned $\Delta_r$.  In Fig.~\ref{fig_dephasing2}, we
rather keep the number of photons in the cavity fixed
($\bar{n}_-=2$) and scan $\Delta_r$.  We see that at large
detunings, the dephasing rate scales as  $\Delta_r^{-2}$.   As a
result, off-resonant irradiation can produce large ac-Stark shifts
($\omega_{\rm ac}$) with minimal dephasing of the qubit.  This can
be used as a single-bit phase gate for quantum computation.

The observed asymmetry in the large $\chi$ case (red solid line of
Fig.~\ref{fig_dephasing2}) is a result of the fact  $\Gm$ depends on
$\bar{n}_-$ and $\bar{n}_+$.  By writing $\bar{n}_+$ as,
\begin{equation}
\bar{n}_+= \bar{n}_- \frac{ \kappa^2/4+(\Delta_r -
\chi)^2}{\kappa^2/4+(\Delta_r + \chi)^2},
\end{equation}
we see that at fixed $\bar{n}_-$, $\bar{n}_+$ can be large for
negative $\Delta_r$. Thus the overall measurement induced dephasing
will be large in that region.

The quality factor for this single qubit gate, $Q$, can be defined
as the coherent phase rotation that can be realized in the total
dephasing time $(T_2^{-1}+\Gamma_m)^{-1}$~\cite{vion:2002}. That is
\begin{equation}
Q = \frac{\omega_{\rm ac}}{2(T_2^{-1}+\Gamma_m)}\approx
\frac{(\Delta_r+\chi)^2}{2\kappa\chi}
\end{equation}
 in the large pull limit and in the ideal situation where dephasing is limited by photon shot noise.
 Moreover, similarly to Rabi oscillations that have been demonstrated experimentally~\cite{wallraff:2005},
 this phase gate could be realized on a time scale which is much faster than $1/\kappa$ since for off-resonant
 irradiation, the cavity is only virtually populated.

To show that the dephasing is minimal during the phase-gate, we have
calculated using~\eq{eq_lineshape_non_gaussian_approx} the linewidth
of the qubit spectrum as a function of input power for the
experimental parameters and a large positive detuning $\Delta_r/2\pi
= 32$ MHz (well away from the peak shown in
Fig.~\ref{fig_dephasing2}).
This is shown in Fig.~\ref{fig_Deph_corrections}  as a blue dashed
line. Here we see that  in the dispersive model, there is no
additional dephasing due to the off-resonant irradiation. That is,
the predicted linewidth stays constant at $\gamma_2$ for the input
powers plotted. At the critical photon number, $n_{\rm crit} =
\Delta^2/4 g^2$  the quality factor for the above experimental
parameters is $17.3$. This quality factor can however be easily
increased by optimizing the system parameters \cite{GatePaper}. For
example at $g/2\pi=100$ MHz and $\Delta/2\pi = 1000$ MHz a quality factor of
$157$ can be reached. (This value is entirely limited by the current
value of $T_2\sim500$ns and not by the direct infidelity of the
phase gate.)
 An advantage of this rf approach over a dc pulse of the
flux or gate charge is that the logical operation can be realized
while biased at the sweet spot.

\subsubsection{Number splitting}
\label{numSec}

\begin{figure}[htb]
\begin{center}
\includegraphics[width=0.45\textwidth]{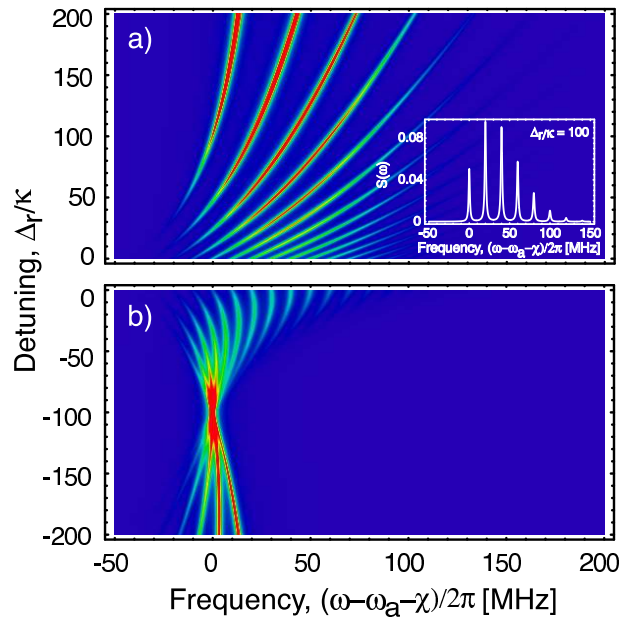}
\caption{\label{fig_density_delta_r} (Color online) Spectrum
$S(\omega)$ as a function of detuning $\Delta_r/\kappa$ at fixed
cavity pull $\chi/\kappa = 100$.  The dephasing rate was set to the
conservative value of $\gamma_2 = 7.6\kappa$.    The average photon
number in the cavity was fixed to $\bar n_-=2$ in panel a)
($\Delta_r \geq 0$) and we choose $\bar n_+=2$ in panel b)
($\Delta_r \leq 0$).  This implies that the measurement beam power
changes with detuning.  Insert: Spectrum at $\Delta_r = \chi =
100\kappa$.  At this detuning, the peaks are split by $2\chi$. More
generally, they are split by $\chi+\Delta$.  Large detuning yields
large splitting but, as shown in Fig.~\ref{fig_dephasing}, this can
be at the expense of small measurable phase shift in the transmitted
field.}
\end{center}
\end{figure}

\begin{figure}[tb]
\begin{center}
\includegraphics[width=.45\textwidth]{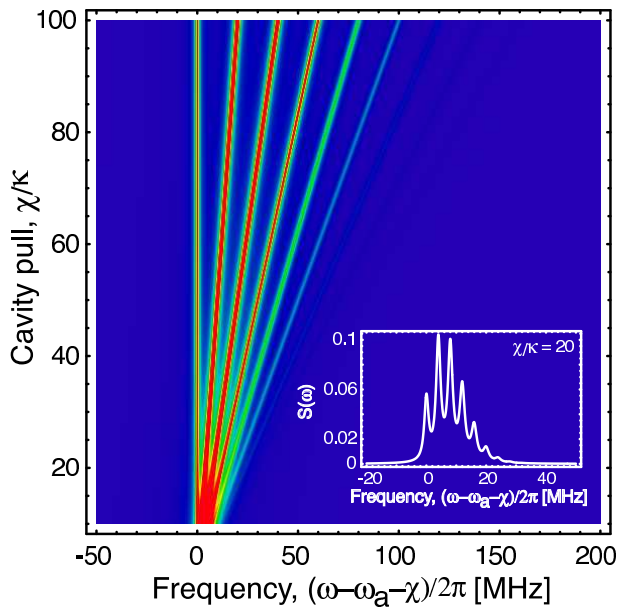}
\caption{\label{fig_density_chi} (Color online)  Spectrum
$S(\omega)$ as a function of the cavity pull $\chi/\kappa$.  The
detuning is $\Delta_r = \chi$ such that the cavity is always driven
at the dressed frequency $\omega_r-\chi$.  Other parameter values
are the same as in Fig.~\ref{fig_density_delta_r}.  Inset:  Spectrum
at $\chi/\kappa  = 20$.  At this experimentally realistic value of
the cavity pull, the number splitting should be resolvable.}
\end{center}
\end{figure}


As explained in the previous sections, the gaussian model and the
P-function approach agree in the small pull limit  $\chi\ll
\kappa/2$.  In the large pull case, the predicted behavior is
however substantially different.  Indeed, when the Lorentzians in
\eq{eq_lineshape_non_gaussian_approx} are separated in frequency by
more than their width, the spectrum $S(\omega)$ will be split into
many peaks with each peak corresponding to a different photon number
in the cavity.  Number splitting was also predicted by Dykman and
Krivoglaz for a different situation, namely an undriven cavity
coupled to a thermal bath \cite{DykKri87}. The number splitting for
our case (driven cavity at zero temperature)
is illustrated in Figures~\ref{fig_density_delta_r} and
\ref{fig_density_chi}.  In Fig.~\ref{fig_density_delta_r}, the
spectrum is shown as a function of frequency and of the detuning
$\Delta_r$ for a fixed $\chi/\kappa=100$.  In
Fig.~\ref{fig_density_delta_r}a) ($\Delta_r > 0$), $\bar{n}_- $ is
fixed and equal to 2 whereas in Fig.~\ref{fig_density_delta_r}b)
($\Delta_r < 0$), $\bar{n}_+=2$. The choice of fixing either $n_+$
or $n_-$ was made so that there is always a small number of photons
in the cavity independent of the state of the qubit. For example,
for $n_-$ fixed,  the number of photons in the cavity when the qubit
is excited, $n_+$, would be small for $\Delta_r > 0$, equal to $n_-$
at $\Delta_r = 0$ and very large at $\Delta_r = -\chi$. The inset
shows a cross-section at $\Delta_r/\kappa = 100$ ($ \Delta_r
=\chi$). For this value of $\Delta_r/\kappa$,  the peaks are very
well separated and the integrated area under each peak obeys Poisson
statistics.  It is interesting to stress that only at $\Delta_r=\chi
$ does the spectrum have a simple Poisson distribution corresponding
to a coherent state with average photon number $\bar n_- =2$. At
$\Delta_r <\chi $ there are more peaks than expected for a coherent
state of this amplitude. Furthermore, at $\Delta_r = -\chi$ the
spectrum is single peaked.

In Fig.~\ref{fig_density_chi}, the spectrum is shown as a function
of the cavity pull $\chi/\kappa$ for a detuning of $\Delta_r= \chi$,
such that the cavity is always driven at the pulled frequency
$\omega_r-\chi$ corresponding to the qubit in the ground state.  In
these plots, we have taken $\gamma_2 = 7.6\kappa$.  Assuming an
experimentally realistic value of $\kappa/2\pi\sim
100$~KHz~\cite{frunzio:2004}, this corresponds to a conservative
$T_2 = 200$~ns~\cite{wallraff:2005}.  As seen on
Fig~\ref{fig_density_chi}, for these parameter values, the peaks
should be resolvable experimentally starting around $\chi/\kappa
\sim 20$ (insert in Fig.~\ref{fig_density_chi}).  Achieving
$\chi/\kappa \sim 20$ in the dispersive limit ($g/\Delta\lesssim
0.1$) requires $g/2\pi\sim20$~MHz.  This value of $g$ was already
realized experimentally~\cite{wallraff:2005} and therefore the
experimental observation of number splitting seems feasible.

\begin{figure}[tb]\begin{center}
\includegraphics[width=.45\textwidth]{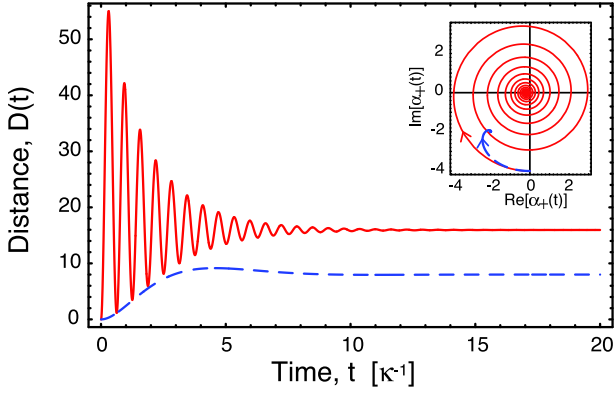}
\caption{\label{fig_distance} (Color online) Distance $D(t) =
|\alpha_+(t)-\alpha_-(t)|^2$ as a function of time for  $\Delta_r  =
\chi$ and $\varepsilon_{\rf} =2\kappa$.  The red (full) line
corresponds to $\chi/\kappa = 5$ and the blue (dashed) line to
$\chi/\kappa = 0.1$.  Other parameter values are the same as in
Fig.~\ref{fig_density_delta_r}.   Inset: Real and imaginary part of
the cavity field $\alpha_+(t)$ as a function of time.  The arrows
indicate direction of time.}
\end{center}
\end{figure}

The behaviour described above can be understood simply as ringing of
a high-Q resonator when its resonance frequency is suddenly changed
when the qubit changes state. Equivalently we can think of this as a
Raman process in which drive photons in the cavity at the time of
the transition are lifted up to the final cavity frequency. As
discussed previously, the calculation of the spectrum  assumes that
the qubit is initially in the ground state with the measurement beam
turned on at a frequency $\omega_d$ detuned by $\Delta_r$ from the
bare resonator frequency $\omega_r$.  In the calculation of the
correlation function $\langle\smm(t)\spp(0)\rangle_{s}$, the qubit
is flipped to the excited state at time $t=0$ and the overlap with
the ground state is calculated at time $t$.  When the qubit is
flipped, the dressed resonator frequency is suddenly changed from
$\omega_r-\chi$ to $\omega_r+\chi$.  Depending on the frequency of
the measurement drive and the quality factor of the cavity, this
sudden change will cause ringing in the cavity.  This is illustrated
in Fig.~\ref{fig_distance} where the distance $D(t) =
|\alpha_+(t)-\alpha_-(t)|^2$ is plotted as a function of time for
two values of $\chi/\kappa$.  For the moderate value of
$\chi/\kappa=5$ (red, full line), the distance is seen to undergo
large oscillations before settling to the steady-state value $D_{\rm
s}$.  For the low $\chi/\kappa$ ratio of 0.1 (blue, dashed line)
there is no ringing due to the abrupt change of dressed cavity
frequency and the distance simply rises to $D_{\rm s}$.

The ringing is also shown in the inset of Fig.~\ref{fig_distance}
where the real and imaginary part of the cavity field $\alpha_+(t)$
are plotted as a function of time, again for $\chi/\kappa=0.1$ and
5.  In the low $Q$ case, as the qubit flips, the cavity field
settles to its new steady state value without large excursions in
the field amplitude, and therefore large changes in photon number.
Only a few different photon numbers contribute and the corresponding
spectrum is single peaked as expected.  In the high $Q$ case, the
field amplitudes performs many cycles  before settling to the
steady-state value.  The cavity therefore probes a large range of
photon numbers and the spectrum shows multiple peaks.  In the time
domain we can see that the qubit correlator, \erf{EQ.Correlation2},
has period recurrences provided $\chi \gg \kappa$. It is these
recurrences in time which give peaks in the spectrum.  These time
domain recurrences were observed in the resonant regime ($\Delta =
\Delta_r = 0$) with Rydberg atoms in
Ref.~\cite{MeuGleMaiAufNogBru05}.

In the case where $\Delta_r = -\chi$ ($\omega_d = \omega_r+\chi$),
the cavity is driven at the dressed cavity frequency corresponding
to the qubit in the excited state.  In this situation, flipping the
qubit does not produce any inelastic Raman scattering (ringing)
since the photons are already at the final cavity frequency.  This
is seen in Fig.~\ref{fig_density_delta_r} at $\Delta_r/\kappa =
-100$ where the spectrum is single peaked.

\begin{figure}[tb]\begin{center}
\includegraphics[width=.45\textwidth]{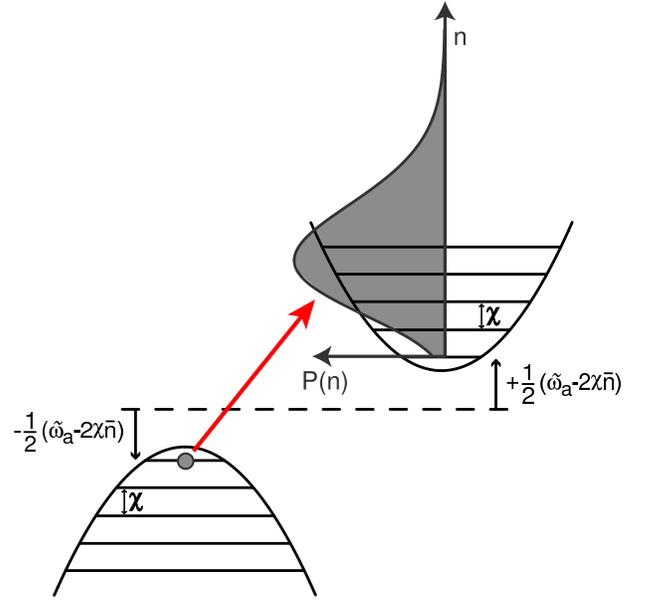}
\caption{\label{Negative_frequencies} (Color online) Pictorial
representation of the dispersive Hamiltonian in a frame defined by
\erf{Uframe} at $\Delta_r = 0$.  When calculating the correlator
$\langle\smm{}(t)\spp{}(0)\rangle$, the qubit is flipped at time
$t=0$.  In the displaced frame, this corresponds to both flipping
the qubit and displacing the oscillator from the vacuum state to a
coherent state with mean photon number $4\bar{n}$.  This
distribution is shown on top of the harmonic oscillator
corresponding to the qubit in its excited state.  This simple
picture explains the observed peaks in the qubit power spectrum and
the presence of negative shifts.}
\end{center}
\end{figure}

Remarkably when the detuning is such that  $\Delta_r<\chi$,  and as
can be seen in Fig.~\ref{fig_density_delta_r}, the spectrum has
peaks at negative frequencies (i.e. below $\tilde\omega_a$) as well
as more peaks than expected. In particular, in the limit that $\chi
\gg \kappa$ and at $\Delta_r = 0$,  the frequencies start at
$\tilde\omega_a - 2\bar{n}_- \chi$ with peak separation $\chi$ and
the spectral weights have a poisson distribution with mean
$4\bar{n}$ and not $\bar{n}$.  To understand the presence of these
peaks, we move the dispersive Hamiltonian~\eq{eq_H_dispersive} to
the frame defined by the unitary operator
\begin{equation} \label{Uframe}
\hat{U}= \hat\Pi_+\hat{D}[\alpha_+]+\hat\Pi_-\hat{D}[\alpha_-].
\end{equation}
Here $\hat{D}[\alpha]$ is the displacement  operator for the cavity
defined by
 \begin{equation} \label{displace}
     \hat{D}[\alpha]=\exp[\alpha \ad-\alpha^*\aop],
 \end{equation}
 and $\hat\Pi_{\pm}$ are the  projectors for the  excited and ground state of the qubit. That is,
we move to a frame that takes both the pointer states out of the
picture. In this frame, ~\eq{eq_H_dispersive}  becomes
\begin{equation}
\hat{H} = \frac{\hbar}{2}[\tilde\omega_a - 2\chi \bar{n}
]\sz{}+\hbar\chi\ad\aop\sz{}.
\end{equation}
Here we have  considered the situation where $\Delta_r = 0$ and
neglected the last term of~\eq{eq_H_dispersive} which only leads to
a small shift of the qubit transition frequency in the present
situation.  This Hamiltonian is  pictorially represented in Fig.
\ref{Negative_frequencies}.  We immediately see that, in this frame,
the qubit transition frequency is reduced by $2\chi\bar n$ from the
lamb shifted frequency $\tilde\omega_\mathrm{a}$.  Moreover, when
the qubit is in the ground state, the Hamiltonian corresponds to a
shifted and inverted harmonic oscillator (LHS of
Fig.~\ref{Negative_frequencies}).  On the other hand, when the qubit
is in the excited state, the harmonic oscillator is shifted but not
inverted (RHS of Fig.~\ref{Negative_frequencies}).  Starting with
the qubit initially in the ground state and the field in a coherent
state of amplitude $\alpha_-$ corresponds, in the frame defined
by~\eq{Uframe},  to a qubit in the ground state and the vacuum state
of the oscillator.  In this frame, flipping the qubit at time $t=0$
corresponds to applying the operator (for $\Delta_r=0$)
\begin{equation}
\hat{U}\dg\spp{}\hat{U}\propto\spp{}\hat{D}[\alpha_--\alpha_+]=
\spp{}\hat{D}[2\sqrt{\bar{n}}].
\end{equation}
The result is to both flip the qubit and to displace the oscillator
to a coherent state of mean photon number $4\bar{n}$.  This is
exactly the observed structure in the qubit power spectrum and each
of the observed peaks corresponds to one of the possible transitions
between these two oscillators. We note that the above is similar to
the Mollow triplet, where transitions both above and below the
atomic transition frequency are possible due to dressing of the
atomic levels by the presence of a strong pump
drive~\cite{mollow:1971}.

Observation of number splitting would constitute a simple test of
number quantization of the field inside the resonator in the
dispersive regime.  In the resonant regime ($\Delta = \Delta_r =
0$), number quantization has been verified in cavity QED using
Rydberg atoms~\cite{brune:1996}.   This was done by looking at the
Fourier components in the probability to find the atom in the
excited state as a function of time.  In the dispersive regime
($\Delta > g$, $\kappa$)  the recurrence time for the cavity field
was too long to be observable \cite{HarocheDis}. In the circuit QED
system it should be possible to reach the strong dispersive limit
$\chi>\kappa,1/T_2$ where it is possible to probe number
quantization. In the dispersive regime the qubit spectrum acts as a
probe of the cavity field.  When the rate at which information about
the cavity state is passed to the qubit faster than the rate at
which the cavity state changes significantly due to damping and
qubit dephasing, it is possible to learn about the statistics of the
field from the qubit spectrum. Note the  above predictions are only
valid in the limit where $\chi^2/\Delta <  \kappa$. When this in not
the case the higher order effects in the Jaynes-Cummings Hamiltonian
[\erf{eq_Hjc}] will become important and will lead to
non-Poissonnian statistics of the resonator field. This will be
discussed further elsewhere. Finally, we note an interesting
proposal by Brune {\it et al.}~\cite{brune:1990,brune:1992} and a
recent experiment~\cite{nogues:1999} to prepare a Fock state of the
cavity field (number squeezing) containing a single photon by
monitoring the state of a continuous beam of atoms sent through the
cavity.

\section{Conclusion}

We have found that due to the ac-Stark shift, quantum noise in the
photon number populating the resonator leads to well characterized
measurement-induced dephasing.  A simple model based on a gaussian
approximation for the phase noise was presented, as well as a more
general model based on the positive P-function.  For the
experimental parameters given in the Letter, both models yield the
same quantitative results which are in very good agreement with the
experiment.  We emphasize that the only adjustable parameter in the
theory is the intrinsic qubit dephasing rate whose only effect is to
give a constant offset to the predicted linewidth.

In the strong dispersive regime, where the cavity pull $\chi$ is
much bigger than the cavity field decay rate $\kappa/2$, the
P-function approach predicts a splitting of the qubit spectrum due
to the discrete quantum nature of the field populating the cavity.
Observation of this prediction would be a confirmation of the
quantized nature of the resonator field. This strong dispersive
coupling regime should be readily achievable with realistic circuit
QED parameters.

\begin{acknowledgments}
We are grateful to Howard Wiseman for discussions. This work was
supported in part by the National Security Agency (NSA) under Army
Research Office (ARO) contract number DAAD19-02-1-0045, the NSF
under Grants No. ITR-0325580 and No. DMR-0342157, and the W. M. Keck
Foundation. AB was partially supported by the Natural Sciences and
Engineering Research Council of Canada (NSERC), the Canadian
Institute for Advanced Research (CIAR) and D-Wave Systems Inc.

\end{acknowledgments}

\appendix
\section{Positive-P representation}
\label{sec_appendix_P}

In this appendix, we show how to solve the master
equation~\eq{eq_master_eq} in the presence of a measurement drive
but take the spectroscopy drive $\veps_\spec=0$.
Substituting~\eq{EQ.StateMatrix} into~\erf{eq_master_eq} yields the
following four coupled differential equations
\begin{align}
&
\begin{split}
\dot{\hat\rho}_{ee} = \:& \kappa {\cal D}[\hat{a}]\hat\rho_{ee}
-\gamma_1 \hat\rho_{ee}
-i\veps_\rf[\hat{a}+\hat{a}\dg,\hat\rho_{ee}]
-i\chi[\hat{a}\dg\hat{a},\hat\rho_{ee}]
\\
&-i\Delta_r[\hat{a}\dg\hat{a},\hat\rho_{ee}],
\end{split}
\label{eq_rho_ee}\\
&
\begin{split}
\dot{\hat\rho}_{gg} = \:& \kappa{\cal D}[\hat{a}]\hat\rho_{gg}
+\gamma_1\hat \rho_{ee}
-i\veps_\rf[\hat{a}+\hat{a}\dg,\hat\rho_{gg}]
+i\chi[\hat{a}\dg\hat{a},\hat\rho_{gg}]
\\
& -i\Delta_r[\hat{a}\dg\hat{a},\hat\rho_{gg}],
\end{split}\\
&
\begin{split}
\dot{\hat\rho}_{eg} = \:& \kappa {\cal D}[\hat{a}]\hat\rho_{eg}
-\gamma_2 \hat\rho_{eg}
-i\veps_\rf[\hat{a}+\hat{a}\dg,\hat\rho_{eg}]
-i\chi\{\hat{a}\dg\hat{a},\hat\rho_{eg}\}
\\
&-i\Delta_r[\hat{a}\dg\hat{a},\hat\rho_{eg}]-i\tilde\omega_a
\hat\rho_{eg},
\end{split}\\
&
\begin{split}
\dot{\hat\rho}_{ge} = \:& \kappa{\cal D}[\hat{a}]\hat\rho_{ge}
-\gamma_2 \hat\rho_{ge}
-i\veps_\rf[\hat{a}+\hat{a}\dg,\hat\rho_{ge}]
+i\chi\{\hat{a}\dg\hat{a},\hat\rho_{ge}\}
\\
&-i\Delta_r[\hat{a}\dg\hat{a},\hat\rho_{ge}]+i\tilde\omega_a\hat
\rho_{ge}.
\end{split}
\label{eq_rho_ge}
\end{align}
Solving these four differential equations would yield a complete
solution. However because of the coupling introduced by $\gamma_1$,
this is not possible analytically for all possible observables.
However for the particular case of computing the dephasing rate, we can
(without error)
 set $\gamma_1=0$ in
 the equations for
 $\hat\rho_{ee}$ and $\hat\rho_{gg}$
while keeping the contribution of relaxation to dephasing in the
equations for the
 off-diagonal components.

To solve Eqs.~\eqq{eq_rho_ee}--\eqq{eq_rho_ge} we express the
density matrix under the positive
P-representation~\cite{walls-milburn}:
\begin{equation}\label{EQ.PFunction}
   \hat \rho_{ij}=\int d^2\alpha \int d^2\beta
    \frac{\ket{\alpha}\bra{\beta^*}}{\bra{\beta^*}\alpha\rangle} P_{ij}(\alpha, \beta).
\end{equation}
Using this expression in Eqs.~\eqq{eq_rho_ee}--\eqq{eq_rho_ge} and
the identities
\begin{eqnarray}
  \hat{a}\ket{\alpha} &=& \alpha \ket{\alpha}, \\
  \hat{a}\dg\ket{\alpha} &=& (\partial_\alpha+\alpha^*/2)\ket{\alpha}, \\
  \bra{\beta^*}\hat{a}\dg &=& \beta\bra{\beta^*}, \\
  \bra{\beta^*}\hat{a} &=& (\partial_\beta+\beta/2)\bra{\beta^*},
\end{eqnarray}
gives four coupled differential equations for the `probability
densities' $P_{ij}$:
\begin{align}
&\begin{split} \dot{P}_{ee} = \:
&  \partial_\alpha [(i\veps_\rf+i\chi \alpha +i\Delta_r \alpha +\kappa \alpha/2)P_{ee}] \\
& +\partial_\beta [(-i\veps_\rf-i\chi \beta -i\Delta_r \beta+\kappa
\beta/2)P_{ee}]
\end{split}\\[0.25cm]
&\begin{split} \dot{P}_{gg} = \:
& \partial_\alpha [(i\veps_\rf-i\chi \alpha +i\Delta_r \alpha +\kappa \alpha/2)P_{gg}] \\
&+\partial_\beta [(-i\veps_\rf+i\chi \beta -i\Delta_r\beta
+\kappa\beta/2)P_{gg}]
\end{split}\\[0.25cm]
&\begin{split} \dot{P}_{eg} = \:
& \partial_\alpha [(i\veps_\rf+i\chi \alpha +i\Delta_r \alpha +\kappa \alpha/2)P_{eg}]\\
&+\partial_\beta [(-i\veps_\rf+i\chi \beta -i\Delta_r \beta +\kappa \beta/2)P_{eg}]\\
&-i2\chi \alpha\beta P_{eg} -\gamma_2 P_{eg}-i\tilde\omega_a P_{eg}
\end{split}\\[0.25cm]
&\begin{split} \dot{P}_{ge} = \:
& \partial_\alpha [(i\veps_\rf-i\chi \alpha +i\Delta_r  \alpha +\kappa \alpha/2)P_{ge}]\\
&+\partial_\beta [(-i\veps_\rf-i\chi \beta -i\Delta_r \beta +\kappa \beta/2)P_{ge}]\\
&+i2\chi \alpha\beta P_{ge}-\gamma_2 P_{ge}+i\tilde\omega_a P_{ge}.
\end{split}
\end{align}
To obtain these expressions, we have assumed that
$P_{ij}(\infty,\infty)=0$ as is usual~\cite{gardiner:2000}.

These equations can be solved simply by making the ansatzen
\be
\begin{split}
P_{ee} = & \delta^{(2)}[\alpha-\alpha_{+}(t)]\delta^{(2)}[\beta-\alpha^{*}_{+}(t)],  \\
P_{gg} = & \delta^{(2)}[\alpha-\alpha_{-}(t)]\delta^{(2)}[\beta-\alpha^{*}_{-}(t)], \\
P_{eg} = & a_{eg}(t)\delta^{(2)}[\alpha-\alpha_{+}(t)]\delta^{(2)}[\beta-\alpha^{*}_{-}(t)], \\
P_{ge} = &
a_{ge}(t)\delta^{(2)}[\alpha-\alpha_{-}(t)]\delta^{(2)}[\beta-\alpha^{*}_{+}(t)]
\end{split}
\ee
and substitute these into each equation. This results in
\begin{align}
\dot{\alpha}_+ =& -i \veps_\rf - i \left(  \Delta_r + \chi  -i \kappa/2\right) \alpha_+ \\
\dot{\alpha}_- =& -i \veps_\rf -i \left( \Delta_r  -\chi - i \kappa/2 \right) \alpha_-\\
\dot{a}_{eg} =&-i\left(\tilde\omega_a - i \gamma_2\right){a}_{eg}
-i2\chi \alpha_+\alpha_-^*{a}_{eg}
\label{eq_a_eg} \\
\dot{a}_{ge} =& i\left(\tilde\omega_a+i\gamma_2\right){a}_{ge}
+i2\chi \alpha_-\alpha_+^*{a}_{ge}.
\end{align}
Solving these simple differential equations completely solves the
master equation~\eq{eq_master_eq} in the absence of mixing due to
$T_1$ effects.   The solution is given in
\erft{EQ.Sol}{eq_alpha_minus}.

\section{Measurement time}
\label{sec_appendix_meas}

In this appendix we show how one can calculate the measurement time
for this system and show how it relates to the quantum
limit~\cite{clerk:2003}. To do this we need to describe how we are
measuring the pointer states (i.e. how the information is
processed). In this experiment, this is done by using heterodyne
detection of the signal that is transmitted from the cavity. In
other words, the full quantum trajectory for the system
is~\cite{wiseman:1993,wiseman:2001}
\begin{equation}
d\rho_{J}(t)=dt{\cal L}\rho_{J}(t)+dt{\cal
H}[(J^*(t)-{\kappa\eta}\langle\hat{a}\dg\rangle)\hat{a}]\rho_{J}(t),
\end{equation} where ${\cal H}[(J^*(t)-{\kappa\eta}\langle\hat{a}\dg\rangle)\hat{a}]\rho_{J}(t)$ is the superoperator representing the non-linear effects of the continuous monitoring and is defined by
\begin{equation}
{\cal
H}[\hat{A}]\rho=\hat{A}\rho+\rho\hat{A}\dg-\langle\hat{A}+\hat{A}\dg\rangle\rho.
\end{equation}
The  measurement record (heterodyne signal) is given by
\begin{equation}\label{measurementrec}
J(t)={\kappa\eta}\langle\hat{a}\rangle+\sqrt{{\kappa\eta}}\zeta(t),
\end{equation} where $\zeta(t)$ is a complex gaussian white noise term, which
is formally defined as
\begin{eqnarray}
{\rm E}[\zeta(t)\zeta(t')]&=&{\rm E}[\zeta(t)]=0\\
{\rm E}[\zeta(t)\zeta^*(t')]&=&\delta(t-t')
\end{eqnarray} where ${\rm E}$ denotes an ensemble average and $\eta=1/[2(N+1)]$ is the inefficiency of the measurement. Here  $N$ is the dark noise and the extra factor of 1/2 is due to the fact that information leaks out of the cavity in both directions and we only monitor transmission~\cite{blais:2004}.

\null From this quantum trajectory  the rate at which information is
obtained about $\langle\hat{a}\rangle$ is $\kappa\eta$. To convert
this to a rate of information gain about
$\langle\hat{\sigma}_{z}\rangle$ we  define the measurement
observable for a time  $\tau$ as
\begin{eqnarray}
{I}(\tau)&=&\frac{1}{\tau}\int_0^\tau  {\rm Re}[J(t)e^{-i\phi}] dt,
\end{eqnarray}
where $\phi$ determines the quadrature in which the information
about the qubit is encoded. This can be determined by
\begin{equation}\label{obs}
\tan\phi =\frac{ {\rm Im} [\alpha_+^s-\alpha_-^s]}{{\rm Re}
[\alpha_+^s-\alpha_-^s]}.
\end{equation}
For example, if $\Delta_r=0$ and $\phi=0$, the information about the
qubit is only encoded into the real part of $\langle\hat{a}\rangle$.
\null From this observable, the mean and variance is
\begin{eqnarray}
\bar{I}(\tau)&=& {\kappa\eta} {\rm Re}[ \langle\hat{a}\rangle e^{-i\phi}] \\
\Delta |I(\tau)|&=&\sqrt{\Big{\langle} [ I(\tau)- \bar{I}(\tau)]^2
\Big{\rangle}} = \sqrt{\frac{\kappa\eta}{2\tau}}.
\end{eqnarray} That is,
if we were to measure the system for a time $\tau$ many times we are
confident that to one standard deviation the value of $\bar{I}$ is
$\kappa\eta{\rm Re}[ \langle\hat{a}\rangle e^{-i\phi}]  \pm
\sqrt{\kappa\eta/2\tau} $.

 If $\tau$ is much shorter than $1/\gamma$ then we can approximate $\bar{I}(\tau)$ with $ \bar{I}_{\pm}(\tau) = \kappa \eta{\rm Re}[ \alpha_\pm^s e^{-i\phi}]$, where the $\pm$ subscript refers to the state of the qubit.
To be able to distinguish between $\langle\hat\sigma_z\rangle=
\pm1$, we require $\Delta\langle\hat\sigma_z\rangle\leq1$ and thus
\begin{equation}
\Delta |I| \leq \frac{|\bar{I} _+  -\bar{I} _-| }{2}=
\frac{\kappa\eta\sqrt{D_{\rm s}}}{2}
\end{equation}  The equality defines the measurement time $t_\mathrm{meas}$.  Using the above, this  can be rewritten as
\be
t_\mathrm{meas} \Gm = \frac{1}{\eta}.
\ee
That is, even for perfect detection efficiency $\eta = 1$, this
approach is a factor of two away from the quantum limit
$t_\mathrm{meas} \Gm = 1/2$ ~\cite{clerk:2003}. This is because even
though we are selecting the correct quadrature in which the
information about the qubit is stored, [using the classical
processing defined in \erf{obs}] we are still measuring the other
quadrature as we are performing heterodyne detection. It is well
known that heterodyne detection measures both the $\phi$ and
$\phi+\pi/2$ quadrature with $1/2$ efficiency
\cite{Haus:1962,yuen:1980}.  Thus if we change the detection scheme
to homodyne detection of the $\phi$ quadrature we can reach the
quantum limit. That is, to reach the quantum limit we require
$\eta=1$ which means we need asymmetric cavities and no dark noise
as well as a detection scheme which extracts only information about
$\hat\sigma_z$.  Note if we did not perform  any classical
processing on the heterodyne signal $J(t)$ then we would be a factor
of four away from the quantum limit.


\end{document}